\documentclass[floatfix,showpacs,amsmath,amssymb,letterpaper,groupaddresses,superscriptaddress]{article}
% ...
\usepackage[caption=false]{subfig}
\usepackage{float}%float images
\usepackage{caption}
\usepackage[english]{babel}
\usepackage[version=0.96]{pgf}
\usepackage{tikz, calc}
\usetikzlibrary{arrows,shapes,snakes,automata,backgrounds,petri,positioning,fit}
\usepackage[latin1]{inputenc}
\usepackage{verbatim}

\usepackage{latexsym}
\usepackage{epsfig}

\usepackage{amsmath}
\usepackage{amssymb}
\usepackage{graphicx}
\usepackage{times}
\usepackage[section]{placeins}
\usepackage{caption}
\usepackage{color}

% ...
\newcommand{\ket}[1]{|{#1}\rangle}

\def\be{\begin{equation}}
\def\ee{\end{equation}}
\def\ba{\begin{eqnarray}}
\def\ea{\end{eqnarray}}

\def\ra{\rangle}

\def\h{\hskip 1cm}

\def\lo{\longrightarrow}

\begin{document}

\vspace{4cm}
\begin{center}{\Large \bf  A simple model of Small-World Quantum Networks}\\
\vspace{2cm}

Ashkan Abedi$\footnote{email:Ashkan@physics.sharif.edu}$ ,\h  Vahid Karimipour$\footnote{email:Vahid@sharif.edu}$\h \\
\vspace{1cm}  Department of Physics, Sharif University of Technology, P.O. Box 11155-9161, Tehran, Iran.\\

\end{center}

\begin{abstract}
A simple model of small world quantum networks, in which a central node plays essential role,  is introduced for sharing entanglement over long distances.  In view of the challenges in setting up advanced quantum labs which allows only few nodes in a network to play a central role, this kind of small world network may be more relevant for entanglement distribution.  It is shown that by only adding a small number of short-cuts, it is possible to produce maximally entangled pairs between arbitrary nodes in the network. Besides this, the threshold values for  the initial amount of entanglement and the distance between nodes, for obtaining a highly entangled states shared between remote points are also investigated.
\end{abstract}

\section{Introduction}
\begin{figure}[H]
	\centering
	\begin{tikzpicture}[
	% node distance=1.5cm
	dot/.style={fill,circle, minimum size=5pt,inner sep=0,node contents={}},
	circ/.style={draw, circle, minimum size=8mm,inner sep=0pt, node contents={}}
	]
	\node [dot, name=n1];
	\node [dot, name=n2, right=of n1];
	\node [dot, name=n3, right=4mm of n2];
	\node [dot, name=n4, right=of n3];
	
	\draw [line width=0.2pt] (n1) -- (n2) (n3) -- (n4);
	
	\node [circ, fit=(n2)(n3), label={[name=R]below:$R$}];
	\node [circ, left, at=(n1.east), name=A];
	\node [circ, right, at=(n4.west), name=B];
	\node at (A |- R) {$A$};
	\node at (B |- R) {$B$};
	\end{tikzpicture}
	\caption{A simple quantum repeater. Circles and dots represent nodes and qubits respectively.}
	\label{fig:repeater}
\end{figure}
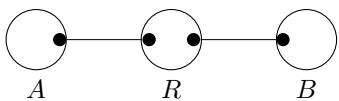

When quantum communication \cite{Qcomm} becomes a reality and a commonplace in the technology of the future, entangled states will be an essential ingredient in the networks which are to carry these communications. One can  imagine that there will be networks of many nodes where  bi-partite maximally entangled states created between distant nodes act in one way or another as carrier of quantum information. The climax of this will be the quantum internet \cite{Qnet}. Being a very fragile resource against noise, which inevitably is more destructive as the distance increases and as time passes, one can use quantum repeaters \cite{repeaters1} \cite{repeaters2} for purifying two partially entangled states shared over short distances to a maximally entangled state over  longer distances. The simplest example is shown in   figure (\ref{fig:repeater}), where one repeater at site $R$  makes a Bell measurement on the two qubits at this site and thus  the partial entanglement \cite{ES} between two pairs $AR$ and $RB$, is swapped with a maximally entangled state at $AB$,
\begin{equation}\label{eq:ro}
	|\phi\ra_{AR}=|\phi\ra_{RB}=\sqrt{\phi}\ket{00}+\sqrt{1-\phi}\ket{11} \h \text{ $ 0\leq \lambda\leq 0.5$}.
\end{equation}

 This measurement produces with probability $ P=2\phi(1-\phi) $ a maximally entangled state between the labs A and B $$|\phi_0\ra_{AB}=\frac{1}{\sqrt{2}}(\ket{00}+\ket{11})$$
and with probability $1-P$ a state $$|\phi_2\ra_{AB}=\frac{1}{\sqrt{\phi^2+(1-\phi)^2}}(\phi\ket{00}+(1-\phi)\ket{11})$$ which is even less entangled than the original one. This state can in turn be converted (distilled) to a maximally entangled state with probability $\frac{2\phi^2}{\phi^2+(1-\phi)^2}$. The total Singlet Conversion Probability \cite{percolation} sums up to 

\be
SCP(2,\phi)=2\phi(1-\phi)+(1-2\phi(1-\phi))\frac{2\phi^2}{\phi^2+(1-\phi)^2}=2\phi.
\ee

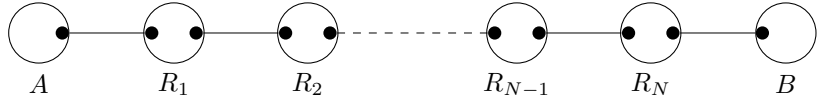
\begin{figure}[H]
	\centering
	\begin{tikzpicture}[
	dot/.style={fill,circle, minimum size=5pt,inner sep=0,node contents={}},
	circ/.style={draw, circle, minimum size=8mm,inner sep=0pt, node contents={}}
	]
	
	\node [dot, name=n1];
	\node [dot, name=n2, right=of n1];
	\node [dot, name=n3, right=4mm of n2];
	\node [dot, name=n4, right=of n3];
	\node [dot, name=n5, right=4mm of n4];
	
	\node [dot, name=n6, right=20mm of n5];
	\node [dot, name=n7, right=4mm of n6];
	\node [dot, name=n8, right=of n7];
	\node [dot, name=n9, right=4mm of n8];
	\node [dot, name=n10, right=of n9];
	
	\draw [line width=0.1pt] (n1) -- (n2) (n3) -- (n4) (n7) -- (n8) (n9) -- (n10);
	\draw [line width=0.1pt, dashed] (n5) -- (n6);
	
	\node [circ, fit=(n2)(n3), label={[name=R]below:$R_1$}];
	\node [circ, fit=(n4)(n5), label={[name=R]below:$R_2$}];
	\node [circ, fit=(n6)(n7), label={[name=R]below:$R_{N-1}$}];
	\node [circ, fit=(n8)(n9), label={[name=R]below:$R_{N}$}];
	\node [circ, left, at=(n1.east), name=A];
	\node [circ, right, at=(n10.west), name=B];
	
	\node at (A |- R) {$A$};
	\node at (B |- R) {$B$};
	\end{tikzpicture}
	\caption{A 1-D network of n quantum repeaters. Circles and dots represent nodes and qubits respectively.}
	\label{fig:n-repeaters}
\end{figure}

For larger distances, i.e. in linear networks like the one shown in  figure (\ref{fig:n-repeaters}), when more repeaters are used, one can repeat this procedure and obtain the total Singlet Conversion Probability. This has been done in \cite{e_distribution} and a very good approximate result is given by

\begin{equation}
SCP(N,\phi)=1-(1-2\phi)\sum_{k=0}^{\lfloor \frac{N}{2} \rfloor} (\phi(1-\phi))^{k} {2k\choose k}.
\label{eq:scp}
\end{equation}
For large $N$, the $SCP(N)$ is bounded above by \cite{e_distribution}
\be
SCP(N,\phi)\lesssim (4\phi(1-\phi))^{\frac{N}{2}},
\ee
which shows that unless we start with perfect Bell pairs ($\phi=\frac{1}{2}$), the success probability for generating Bell pairs between distant labs decreases exponentially by the number of repeaters. This discouraging result for linear networks, has prompted investigation of entanglement percolation \cite{percolation}  and entanglement distribution \cite{e_distribution} in regular one- and two-dimensional networks with various  (i.e rectangular, triangular and hexagonal) geometries.  In particular it has been shown that the SCP can reach  unit value in a finite number of steps, provided that the initial entanglement between pairs exceeds a threshold value $\phi_t$, which depends on the geometry. To come a little closer to networks in real life, quite recently random quantum networks have been studied \cite{q_random}. In these networks, which are adaptation of the  Erdos and Renyi random graphs \cite{random} to the quantum domain,  pairs of nodes are maximally entangled with a probability $p$ and are disentangled with probability $1-p$. It has been shown that any quantum subgraph can be generated by Local Operation and Classical Communication (LOCC) if $p\sim N^{-2}$, $N$ being the size of the nodes.\\
 
% Different approaches has been taken to study quantum random networks.  
While random graphs have some interesting features in common with real networks, like short average distance between nodes, they fail to simulate all the properties of real networks.  
In particular they do  not show the important property of clustering. In real networks, two nodes which are the neighbors of a given node have a much higher probability of being connected together, compared with two arbitrary nodes.  This is measured by the clustering coefficient C \cite{book2} which is the the fraction of neighbors of a given site which are nearest neighbors of each other. For a rectangular graph in any dimension $C=0$, and for a fully connected graph, $C=1$. For a regular 2D triangular lattice, $C=\frac{2}{5}$ (any node in such a graph has 6 neighbors which are connected to each other, while the total number of pairs is ${6\choose 2}=15.$). Small world networks \cite{Watts} are models of networks in which both properties, namely small average distance and clustering property are present. \\

\begin{figure}[h]
	\centering
	\includegraphics[scale=0.5]{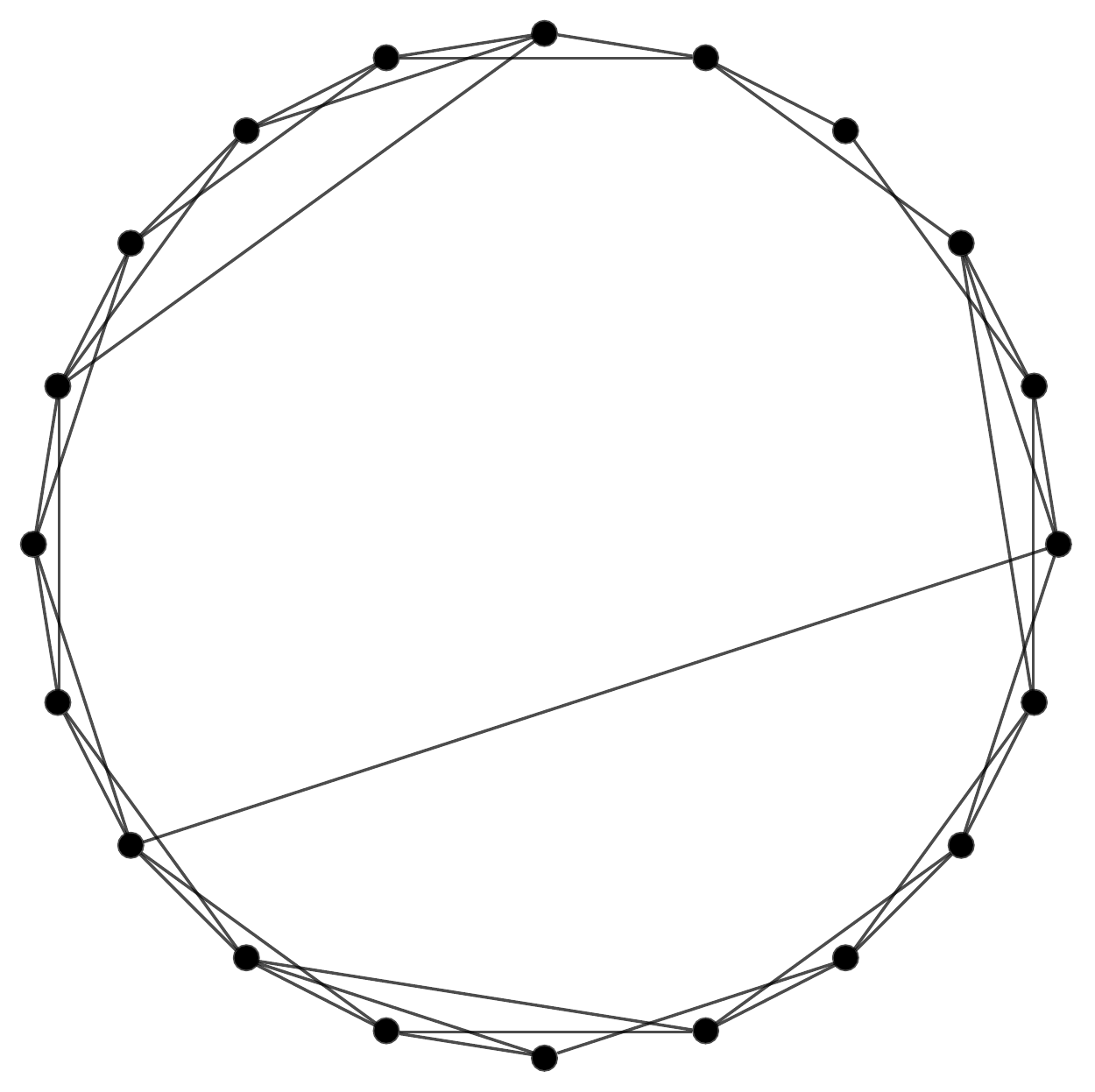}
	\caption{A simple model of small-world network model with 20 nodes introduced by Watts and Strogatz when $s=4$ and $p=0.1$ in this network.}
	\label{fig:watts1}
\end{figure}

The ubiquity of small world networks makes it very desirable to investigate the problem of entanglement generation in such networks.  A simple model was first introduced by Watts and Strogatz in \cite{Watts}, (figure (\ref{fig:watts1})), where a ring of $N$ sites is considered in which every site has a number of $s$ neighbors. Depending on the value $s$, this provides the clustering property of the network.
The other element of a real network is the low average distance between the nodes which is provided by random short-cuts in the lattice. Therefore in figure (\ref{fig:watts1}), with probability $p$ a node's connection to its nearest neighbor is cut in favor of a new long range connection to a distant node. Small world networks have been studied in the quantum domain in a number of contexts, like transportation of excitations\cite{muel1, muel2}, and also as realistic models of random networks for entanglement distributions\cite{q_random, e_distribution}.

\begin{figure}[h]
	\centering
	\includegraphics[scale=0.18]{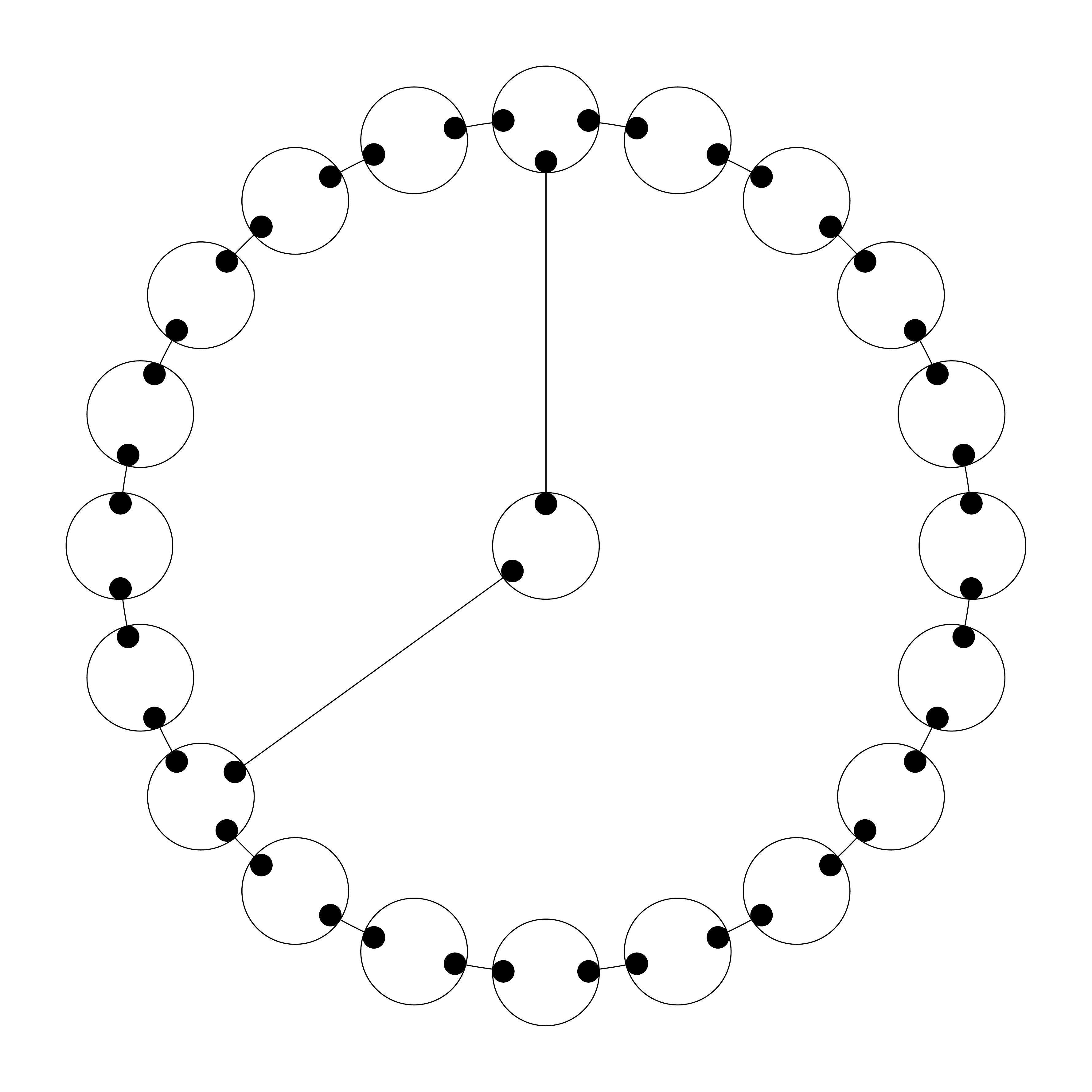}
	\caption{A quantum small-world network with 21 nodes. Circles and dots represent nodes and qubits respectively. $n=20, m=2$ and $p=0.1$ in this network.}
	\label{fig:swn_example}
\end{figure}

In this work, we are not concerned the Watts and Strogaz whose quantum version has been studied in \cite{q_random}. Instead we consider an even simpler model first introduced in \cite{Dorogovtsev}, where all shortcuts are made, with a probability $p$ to a central node in the graph. 
This model is much more relevant to quantum networks since it is plausible that there should be only very few nodes (labs) which can afford full-fledged facilities for producing entangled states and sharing it with arbitrary labs in a future quantum network, figure (\ref{fig:swn_example}). The central node in this figure can even represent a satellite to which other labs on the Earth which are out of sight of each other, share entangled states. The shared entangled states with the satellite are depicted by shortcuts to the center in figure (\ref{fig:swn_example}). Entanglement swapping on the satellite then  provides entangled states between these remote labs.  Under these circumstances, natural questions arise which we will investigate in this paper. In particular we are interested in the following questions: \\

1-How the threshold value of the initial entanglement between the pairs depend on the number of shortcuts, if we want to maximally entangle  distant points in this network?\\

2- For a given value of initial entanglement between the pairs, is there a critical number of shortcuts above which arbitrary pairs of labs far from each other can still establish  maximally entangled states between themselves? \\

The answers to these questions are summarized in figures (\ref{fig:scp_n_2}) and (\ref{fig:r0}). From figure (\ref{fig:scp_n_2}) we see  that even a small number of shortcuts and a small value for clustering coefficient has an appreciable effect on entanglement distribution in these networks. \\

The paper is  structured: In  section (\ref{small}), we briefly review the small network model of \cite{Dorogovtsev} and derive its properties. In section (\ref{mysmall}) we calculate the singlet conversion probability for this network and present the results in two figures (\ref{fig:scp_n_2}) and (\ref{fig:r0}).   The paper ends with a discussion.

% Quantum small-world network

\section{A simple model of small world network}\label{small}
Figure (\ref{fig:directed1})  shows a  simple model of small world network first introduced in  \cite{Dorogovtsev}. There is one node (central lab) in the network which has a central role and other nodes in the graph are connected to this node with probability $p$, creating shortcuts which drastically lower the shortest path between remote points which is the characteristic property of small world networks.  Due to its simple structure, many of the statistical properties of these networks have already been studied in various papers \cite{SModels}. Here we want to study the same network in the context of quantum technology. Thus in the network shown in figure (\ref{fig:swn_example}), each link denotes a partially entangled state of the form (\ref{eq:ro}) shared between the two nodes (labs) at the end of the link. The structure of the network indicates that each lab can share these states with its nearest neighbor lab, but the central node can share partially entangled states (\ref{eq:ro}) with any other lab. \\  

   Let $n$ be the total number of nodes on the circle, then the average number of shortcuts is  given by $m=np$. When $n\lo \infty$, two distinct limits can be considered, one in which $p\lo 0$ such that the number of shortcuts $m$ remains finite and the other in which $p$ is finite in which case the number of shortcuts too tends to infinity.  The clustering coefficient of this network is $p^2$, since the probability that the two neighbors of a given node are connected to the central node is $p^2$.   For simplicity we count the length of each shortcut as $\frac{1}{2}$. Consider two points whose shortest path along the ring is $r$. We call this the {\bf regular distance} of the two nodes.  In the absence of shortcuts, this is the only shortest path between the two points. When $p>0$, the average shortest path between these points is denoted by $\overline{l}(r)$. This is called the {\bf actual distance} of the two nodes. We now consider two different networks, one in which the links on the ring are directed and one in which there is no direction on these links. In both cases the shortcuts to the ring are not directed. Obviously analysis of the directed ring is much easier than the undirected ring. However in the context of quantum networks, where a link means the possibility of creating an entangled pair, there is no point in assuming a direction on the links and it is the undirected graph which we should take into account. Nevertheless it is instructive to first review some of the classical properties of the directed rings for entanglement distribution and then extend this study to the undirected graphs.

\subsection{Small world graph with directed links}
Consider the directed graph in figure (\ref{fig:directed1}). Let $a$ and $b$ be two nodes with regular distance $r$. The probability that their actual distance is $\ell$ is given by \cite{Dorogovtsev} 
\ba
P(\ell|r)&=&\ell p^2(1-p)^{\ell-1}, \h \ell<r\cr 
P(r|r)&=&(1-p)^{r+1}.
\ea
 The average shortest path, between any two points can be calculated in closed form, although the expression is not so illuminating. Instead we draw in {figure (\ref{fig:lbar}) the average shortest path ($\bar{l}$) as a function of $p$ for several values of $N$. 

\begin{figure}[h]
	\centering
	\includegraphics[scale=0.2]{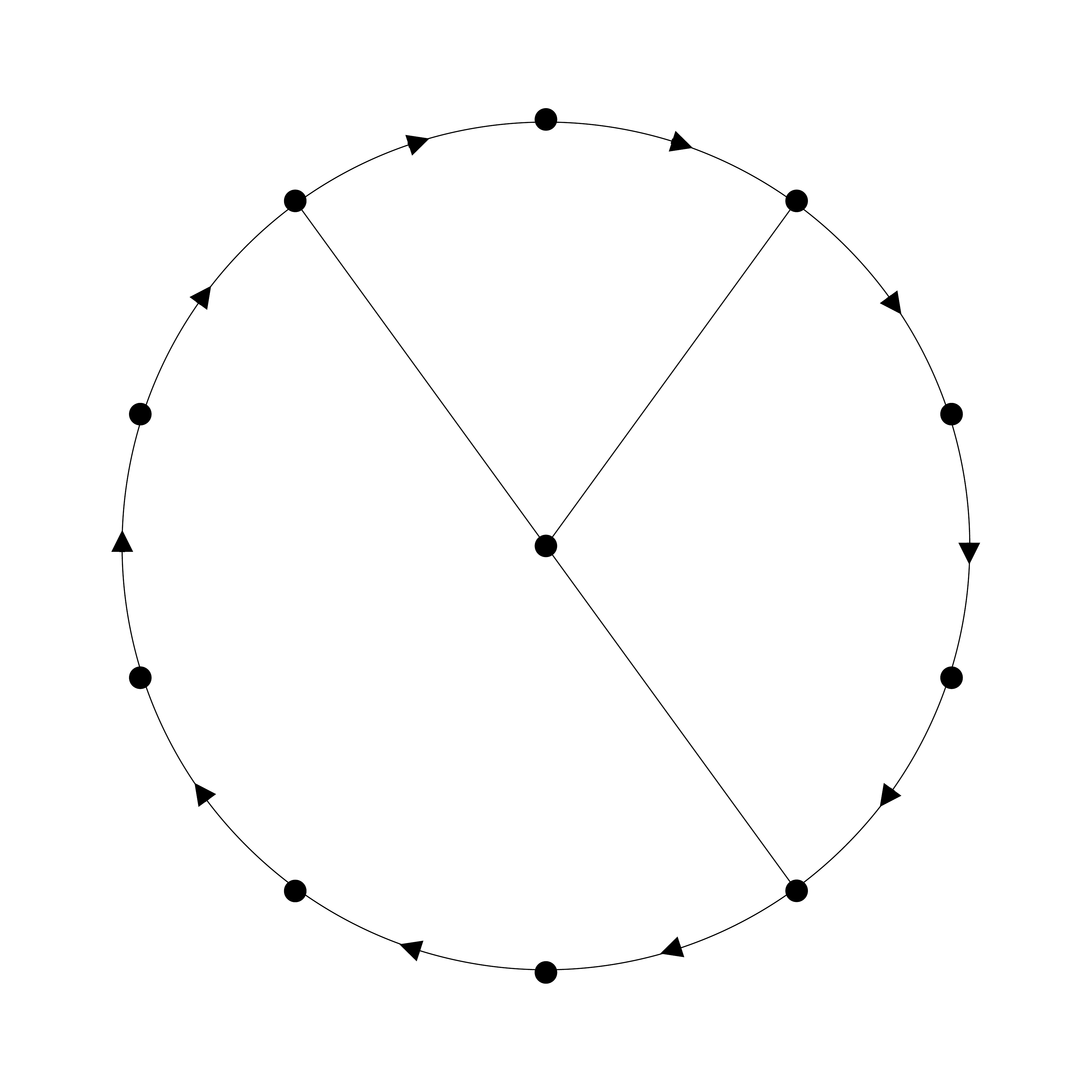}
	\caption{Simple model of small world graph with directed links.}
	\label{fig:directed1}
\end{figure}

\begin{figure}[h]
	\centering
	\includegraphics[scale=0.5]{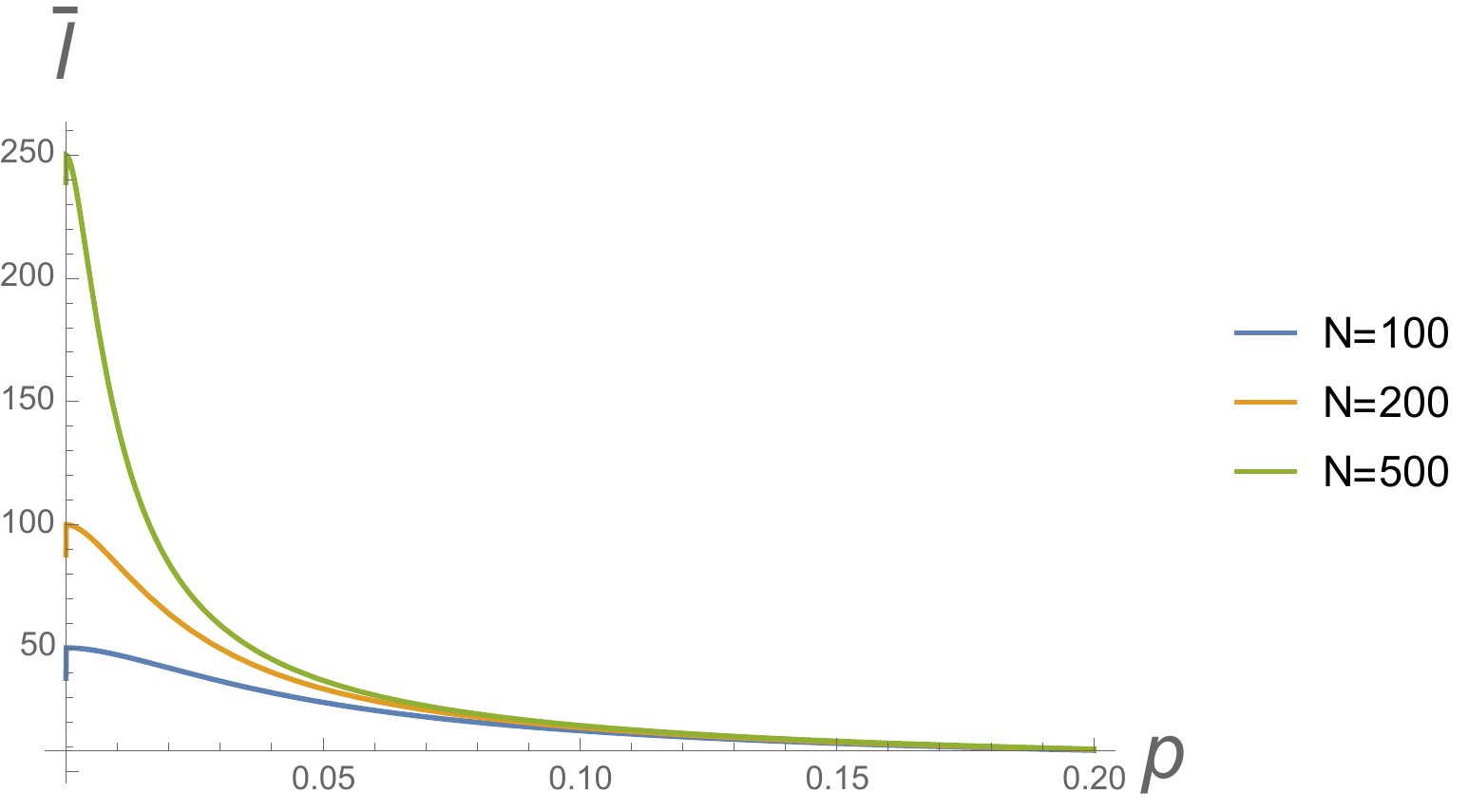}
	\caption{Average shortest path ($\bar{l}$) with respect to $p$ in a directed small world network for $N=100,200 \text{ and } 500$.}
	\label{fig:lbar}
\end{figure}

When the graph is undirected, analytical calculation of the probabilities and the average length become more involved. The result is \cite{Dorogovtsev}

  \ba
  P(\ell|r)&=&p^2(1-p)^{2\ell-4}(2-p)(2\ell-p\ell-2), \h 1<\ell<r\cr
  P(\ell|r)&=&p^2, \h \ell=1\cr
  P(r|r)&=&1-\sum_{\ell=1}^{r-1} p(\ell|r).
  \label{eq:p_undirected}
  \ea
  However the qualitative behavior of the probabilities do not have much difference with the directed network. 
 \section{Entanglement generation in a small  world quantum network}\label{mysmall}
We are now in a position to consider the problem of entanglement generation
 in the small world network. Let $a$ and $b$ be two points with regular distance $r$. Obviously we should consider only undirected graphs for which $P(\ell|r)$ is given in equation (\ref{eq:p_undirected}). In the absence of any shortcuts, when $p=0$ the singlet conversion probability for these two sites is given by $SCP(r,\phi)$ as in equation (\ref{eq:scp}). When $p\ne 0$, the two sites can create a singlet with probability through the shortest paths provided for them. The average number of shortcuts in this case is equal to $m=np$. Each path is provided with probability $P(\ell|r)$ and the average $SCP$ for these two sites will be given by 

\be
\overline{SCP}(r,\phi,m)=\sum_{\ell=1}^r SCP(\ell,\phi)P(\ell|r),
\ee 
where $SCP(\ell,\phi)$ is given in (\ref{eq:scp}) and $P(\ell|r)$ is given in (\ref{eq:p_undirected}).  To answer the first question posed in the introduction, i.e. the probability of success for creating a singlet between the two distant labs, we consider
$
SCP(r, \phi,m):=\overline{SCP}(\frac{n}{2},\phi,m)$ for three different distances $r=20, 80$ and $500$ in a network of $N=1000$ nodes. Several interesting features are observed in these figures.  First we see that when the initial shared states are maximally entangled (i.e. $\phi=0.5$), there is no need for any shortcuts in order to entangle distant nodes, no matter how far apart they are. This is because these nodes can use the links along the circular network and can successfully extract a Bell state with SCP=1. No matter what is the distance between these labs, the SCP is always 1 in view of Eq. (\ref{eq:scp}). For short distances say $r=20$ (figure 7.a), one can still obtain large SCP, with a small number of shortcuts,  even if the shared states are not maximally entangled. For any value of $\phi<0.5$, there is a threshold number of shortcuts above which one can establish maximally entangled states with high SCP. This threshold value increases with distance, but it is seen from the figures (7.b) and (7.c) for large distances, its value approaches a fixed value. This is explicitly depicted in figure (\ref{fig:scp_n_2}) . It is seen that for short distances, the distant labs can always use the links around the circular network to extract maximally entangled states. At these distances the addition of shortcuts does not have much effect. As the distance increases, we find a threshold distance beyond which, a certain number of shortcuts, is needed for attaining an SCP of 2/3. For the value of $\phi=0.45$, the threshold distance is around $r_0\approx 20$ and the threshold value is around $m\approx 50$. If we now  increase the threshold SCP to $\frac{3}{4}$, the required number of shortcuts  increases to twice the previous value and also the distance below which an SCP of $\frac{3}{4}$ is possible in the absence of shortcuts reduces to half of its previous value. These are shown in figure (\ref{fig:r0}). The interesting feature is that the number of shortcuts remains constant after a distance. The reason is that by increasing the number of shortcuts and making them denser in the network, the shortest distance between two points does not necessarily decrease further.

\begin{figure}[h]
	\subfloat[$r=20$\label{sfig:43}]{%
		\includegraphics[width=0.32\linewidth]{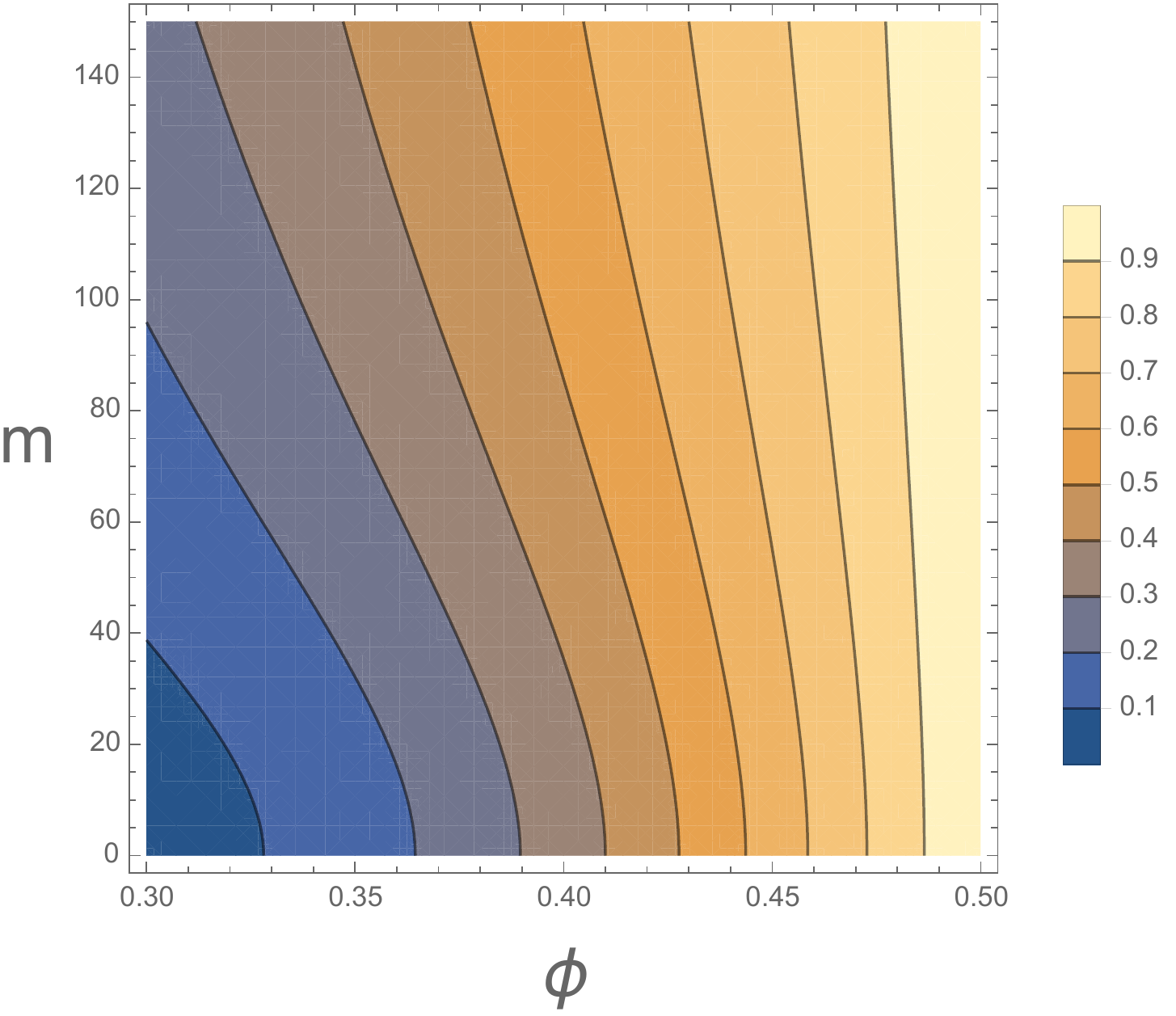}%
	}\hfill
	\subfloat[$r=80$\label{sfig:45}]{%
		\includegraphics[width=0.32\linewidth]{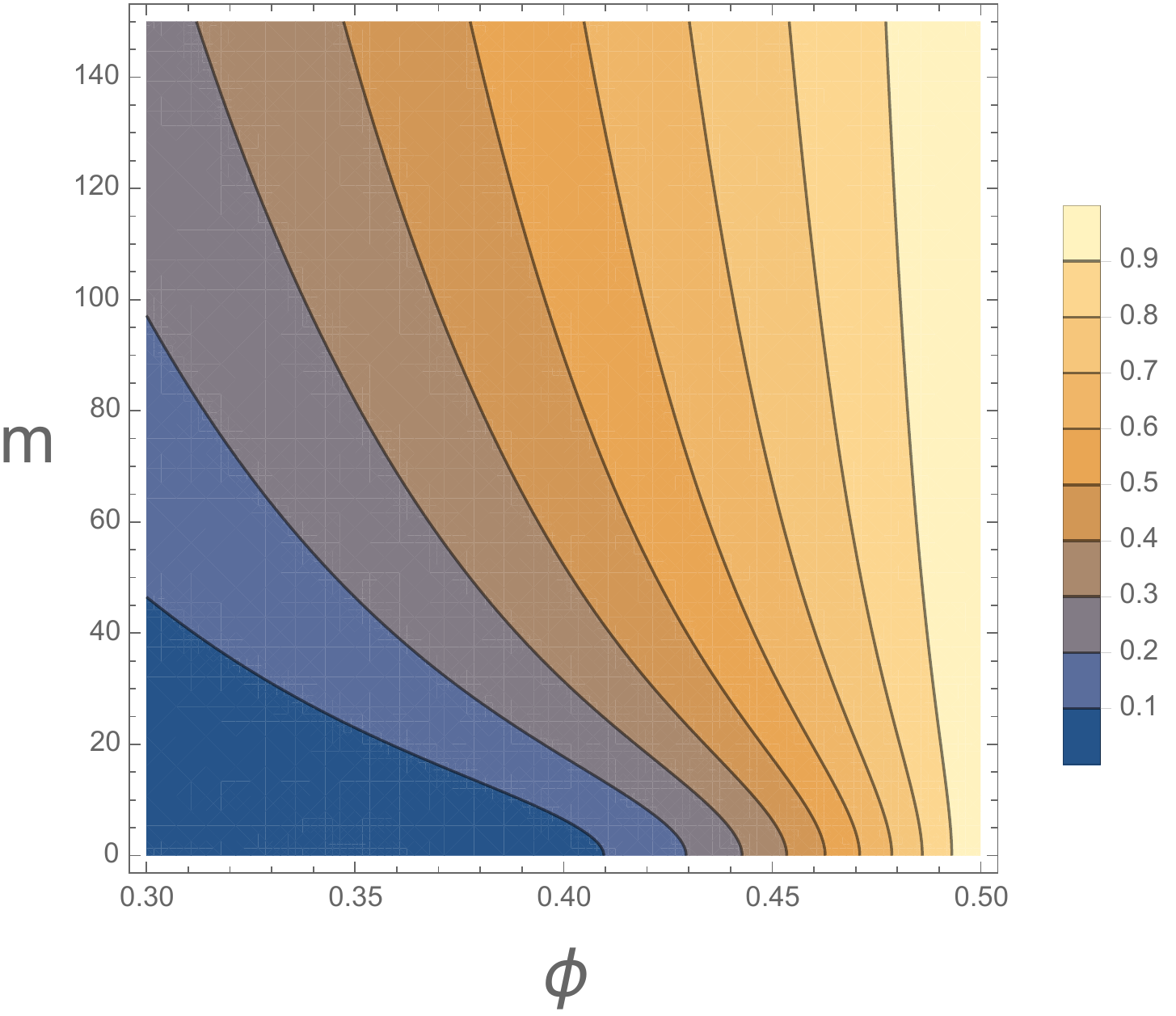}%
	}\hfill
	\subfloat[$r=500$\label{sfig:47}]{%
		\includegraphics[width=.32\linewidth]{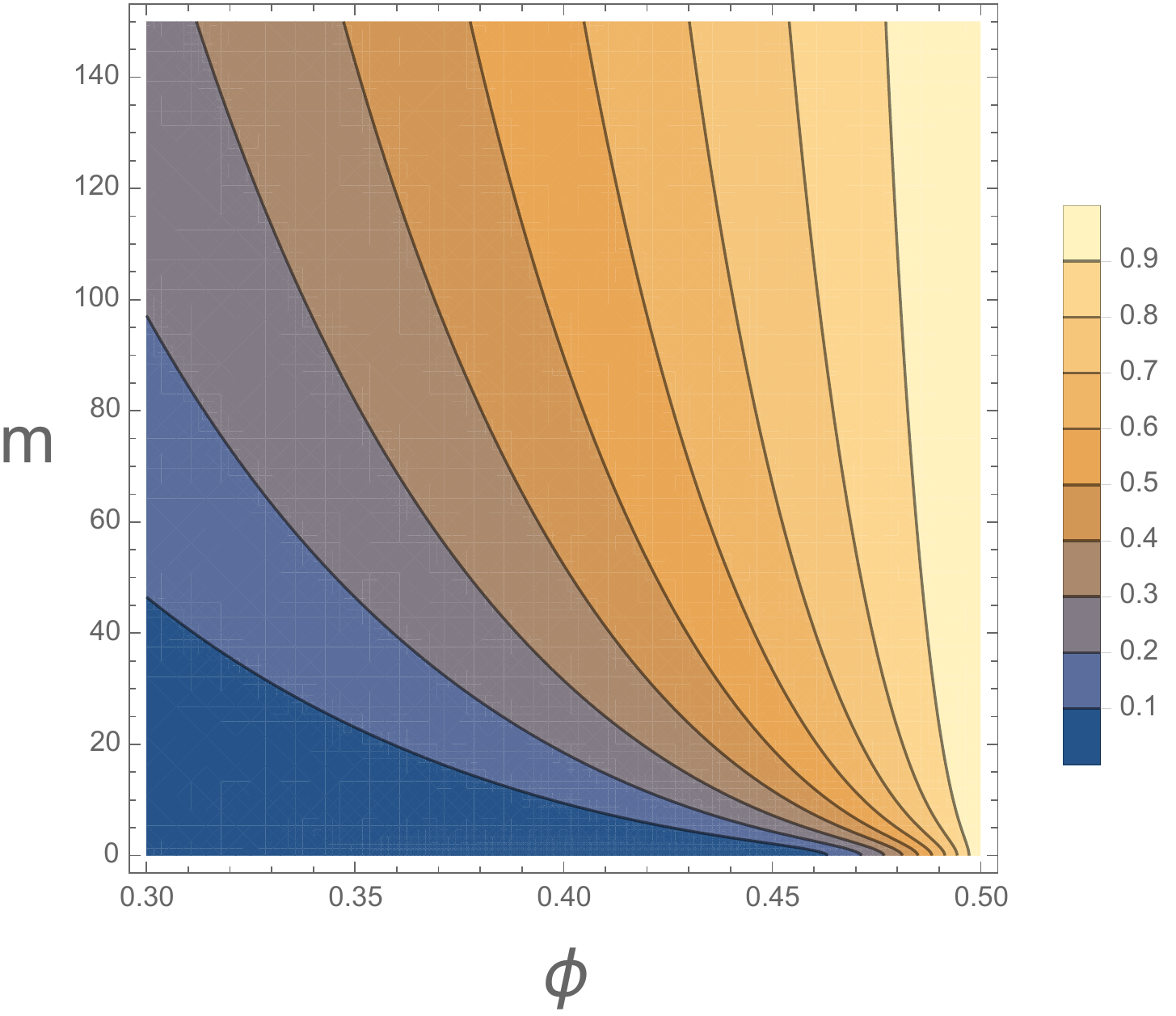}%
	}\hfill
	\caption{The singlet conversion probability (indicated by color) for pairs of points at three different distances (r=20, 80 and 500) for a network of $N=1000$ nodes. The plots show dependence of SCP on the amount of initial entanglement and the number of shortcuts.  }
	\label{fig:scp_n_2}
\end{figure}

\begin{figure}[h]
	\subfloat[$
	SCP=2/3,\ \ \phi=0.45$\label{sfig:43}]{%
		\includegraphics[width=0.38\linewidth]{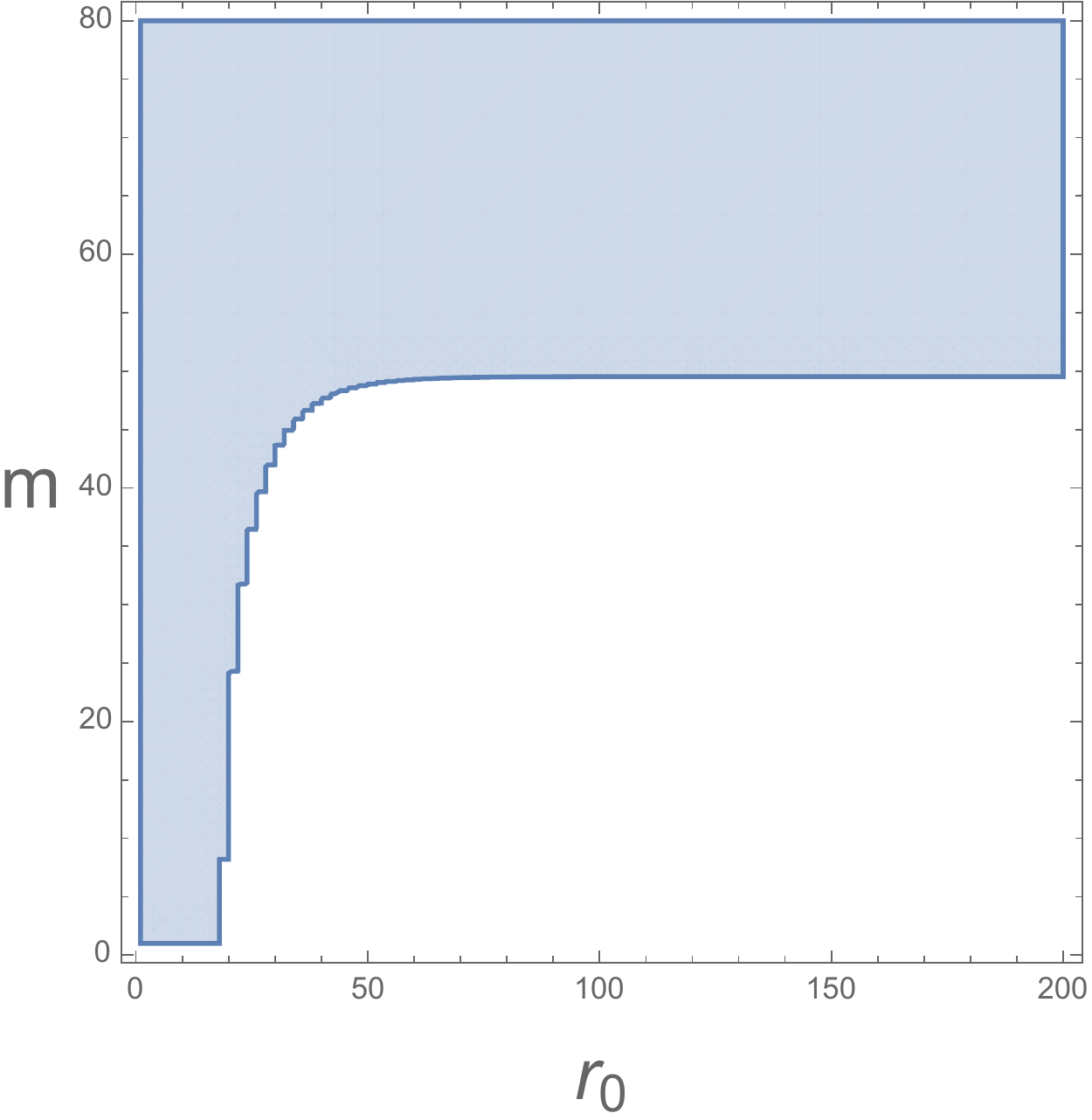}%
	}\hfill
	\subfloat[$SCP=3/4,\ \ \phi=0.45$\label{sfig:45}]{%
		\includegraphics[width=0.38\linewidth]{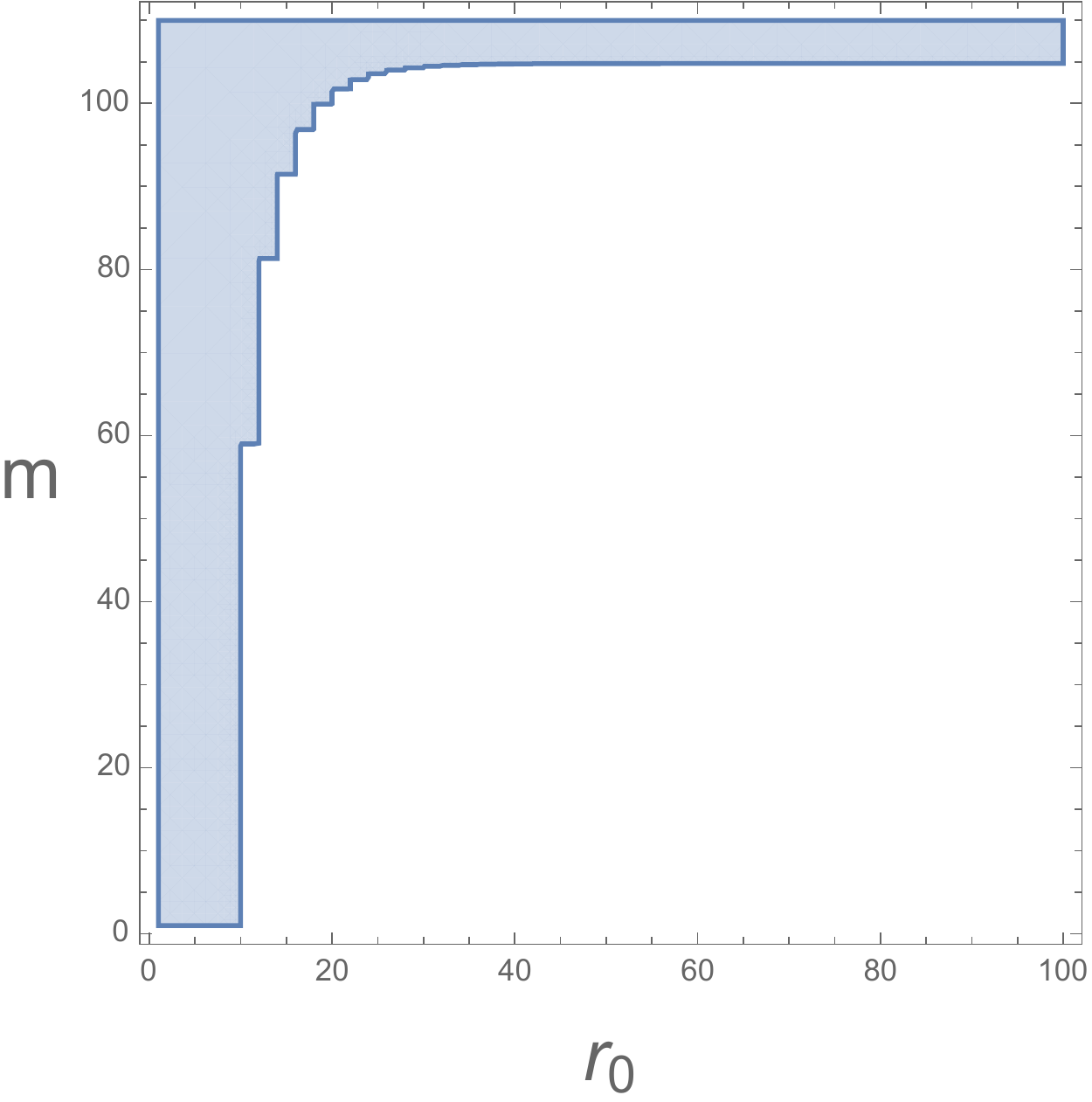}%
	}\hfill
	\caption{a) The area in which the threshold is bigger than 2/3 is shown by blue color. The initial value of entanglement is  $\phi=0.45$. If the distance between two labs is less than $r_0\approx 20$, they can use the links around the circular network to extract maximally entangled states. At these distances the addition of shortcuts does not have much effect. Beyond this threshold distance, a certain number of shortcuts around 50, is needed for attaining an SCP of 2/3. b) For the same value of initial entanglement $\phi=0.45$, but an SCP of 3/4. The blue area has shrunk. Here for distance bigger than 10 links, shortcuts are necessary in order to achieve an SCP of 3/4. In both figures, the  interesting feature is that the number of required shortcuts saturates to a constant value after a specific distance. The reason is that by increasing the number of shortcuts and making them denser in the network, the shortest distance between two points does not necessarily decrease further. This result has certainly practical consequences.   }
	\label{fig:r0}
\end{figure}

\section{Conclusion}
  In this letter we have briefly discussed a very simple model for entanglement distribution in small world networks. The network has a central node ( a central lab, i.e. a satellite, which can share entangled states with all the other labs in the network. We have determined to what extent distant labs in the network can share maximally entangled states, if initially they have shared only partially entangled states with their neighboring labs and with some probability with the central lab. We have obtained the threshold values for the number of shortcuts, the initial value of entanglement and the distance between nodes, for obtaining a highly entangled states shared between remote points. This last quantity is measured by a quantity called Singlet Conversion Probability (SCP). The results are shown in figures (\ref{fig:scp_n_2}) and (\ref{fig:r0}). It would be interesting to investigate how the number of central labs can change this results. Moreover the same model with one central node, can be modified to exhibit one other characteristics of small world networks, namely the clustering property. This is done simply by connecting each node on the ring, not only to its nearest neighbors, but to a small number of its neighbors. We have found that this clustering property does not  appreciably affect the main results reported in figures (\ref{fig:scp_n_2}) and (\ref{fig:r0}).

\section{Acknowledgements} 
This research was partially supported by a grant no. 96011347 from the Iran National Science Foundation. The work of V. K. was also partially supported by a grant from the research grant system of Sharif Univeristy of Technology.  .

\end{document}